\documentclass[12pt]{article}
\usepackage{amsmath,amsfonts,amssymb,amsthm}
\newtheorem{thm}{Theorem}[section]
\newtheorem{pro}[thm]{Proposition}

\theoremstyle{definition}

\numberwithin{equation}{section}
\newcommand{\CA}{\mathcal{A}}
\newcommand{\CH}{\mathcal{H}}

\newcommand{\CK}{\mathcal{K}}
\newcommand{\CV}{\mathcal{V}}

\newcommand{\Z}{\mathbb{Z}}
\newcommand{\R}{\mathbb{R}}

\newcommand{\ket}[1]{|#1\rangle}

\newcommand{\acts}{\triangleright}

\newcommand{\oh}{{\tfrac{1}{2}}}

\newcommand{\ts}{\otimes}

\newcommand{\sq}{\unskip\nobreak\kern5pt\nobreak\vrule height4pt width4pt depth0pt}   
\newcommand{\su}{\mathfrak{su}}    

\title{Quasi-Dirac Operators and Quasi-Fermions}
\author{Andrzej Sitarz
\thanks{Partially supported by MNII Grant 115/E-343/SPB/6.PR UE/DIE 50/2005--2008} \\
Institute of Physics, Jagiellonian University, \\
Reymonta 4, 30-059 Krak\'ow, Poland}
\begin{document}
\maketitle
\begin{abstract}
\noindent
We investigate examples of quasi-spectral triples over two-dimensional
commutative sphere, which are obtained by modifying the order-one
condition. We find equivariant quasi-Dirac operators and prove that
they are in a topologically distinct sector than the standard Dirac
operator.
\end{abstract}


\section{Introduction}
Noncommutative Geometry offers a new insight into classical differential
geometry of spin manifolds. The Connes' Reconstruction Theorem, which
under certain assumptions is proved in \cite{BFV-Book} and has been
recently extended \cite{Connes08} provides the equivalence between the
spectral triples for commutative geometries and the geometry of spin
manifolds.

In this note we investigate what happens if one relaxes one of the axioms
of the spectral triple construction by allowing the order-one condition
to be satisfied up to compact operators. This step is motivated by strictly
noncommutative $q$-deformed situation. In the case of $SU_q(2)$ \cite{suq2,suq2-i}
and the Podle\'s equatorial quantum sphere \cite{s2q-eq} it was found that the
order one condition and the commutant condition can be satisfied only up to
compact operators. It has been conjectured that for commutative geometries
similar weakening of this axiom shall not lead to any significant change apart
from small (compact) perturbation of the Dirac operator. In our example we show that
this is not true, in particular, the spectral geometries obtained in this way are
topologically distinct.

For the two-dimensional sphere we determine a class of examples of equivariant quasi-spectral
geometries. Using the local index calculations we prove that they pair differently with an element
of K-theory and therefore are mutually inequivalent.

The two-dimensional commutative sphere has been a favorite toy-model for description of noncommutative
geometry but rather in a local (coordinate) description \cite{BFV-Book, DaSilva}. A {\em global}
approach was first used by Paschke \cite{Paschke-Th} and could also be easily obtained by taking
$q=1$ limit of the algebraic description of Podle\'s spheres. Our work is partially motivated by
an attempt to make sense of the classical limits of spectral geometries obtained for $q$-deformed
spheres in \cite{Sitarz-quasi}.

\section{Spectral Geometry of the Sphere}

We follow here the construction of equivariant spectral geometry
as described in \cite{Sitarz-eq}. The symmetry, which we use is
the enveloping algebra of $\su(2)$, with generators $L^3, L^\pm$
satisfying:
$$ [L^\pm, L^3] = \mp L^\pm, \;\;\;\;\; [L^+, L^-] = 2 L^3. $$
We work with the commutative star algebra $\CA(S^2)$ of polynomials
in $A^*,B,B^*$, satisfying the radius relation:
\begin{equation}
A^2 + B B^* = 1. \label{radius}
\end{equation}
The generators of the polynomial algebra $A,B,B^*$ are the spherical
harmonics of degree $1$. The action of the $\su(2)$ Lie algebra
on the generators is:
\begin{flalign*}
& L^+ \acts B = 0, &
& L^- \acts B = - 2A, &
& L^3 \acts B = B, \\
& L^+ \acts B^* = 2 A,&
& L^- \acts B^* = 0, &
& L^3 \acts B^* = - B^*, \\
& L^+ \acts A = -B, &
& L^- \acts A = B^*, &
& L^3 \acts A = 0. \\
\end{flalign*}
\subsection{The equivariant representations}

Let $V_l$, $l=0,\oh,1,\ldots...$ denote the $2l\!+\!1$-dimensional
representation of the Lie algebra $\su(2)$. The orthonormal
basis of each $V_l$ shall be denoted by $\ket{l,m}$,
$m=-l,-l+1,\ldots,l-1,l$. The representation $\rho$ of $\su(2)$
on $V_l$ is:
$$
\begin{aligned}
L^+ \ket{l,m} &= \sqrt{l-m}\sqrt{l+m+1} \ket{l,m+1}, \\
L^+ \ket{l,m} &= \sqrt{l+m}\sqrt{l-m+1} \ket{l,m-1}, \\
L^3 \ket{l,m} &= m \ket{l,m}.
\end{aligned}
$$
We recall that for any algebra $\CA$ on which a Lie
algebra $\mathfrak{l}$ acts by derivation, the representation
$\pi$ is $\mathfrak{l}$-equivariant if there exists
a representation $\rho$ of $\mathfrak{l}$ such that:
\begin{equation}
\rho(\ell) \pi(a) = \pi(\ell \acts a) + \pi(a) \rho(\ell),
\label{equiv}
\end{equation}
holds for all $a \in \CA$, $\ell \in \mathfrak{l}$ on the linear
space $\CV$.

Since we know the representation theory of $\su(2)$ we shall
decompose $\CV$ as a direct product of irreducible
representations of $\su(2)$:
$$ \CV = \bigoplus_l V_l.$$
We have:
\begin{pro}
For each $N=0,\pm\oh,\pm1,\pm\frac{3}{2},\ldots$ there
exists an irreducible unitary representation $\pi_N$ of
$\CA(S^2)$ on the Hilbert space $\CH_N$, which is the
completion of $\CV_N$:
$$ \CV_N = \bigoplus_{l=|N|}^\infty V_l,$$
which is equivariant on $\CV_N$ with respect to $\su(2)$ action.
The explicit form for $\pi_n$ on the orthonormal basis
$\ket{l,m} \in V_l$, is:
\begin{equation}
\begin{aligned}
\pi_N(B) \ket{l,m} =& \sqrt{(l+m+1)(l+m+2)} \,\alpha^+_l \, \ket{l+1,m+1} \\
&+\sqrt{(l+m+1)(l-m)} \,\alpha^0_l \, \ket{l,m+1}\\
&+ \sqrt{(l-m)(l-m-1)} \,\alpha^+_{l-1} \,\ket{l-1,m+1},
\label{rep-B}
\end{aligned}
\end{equation}

\begin{equation}
\begin{aligned}
\pi_N(B^*) \ket{l,m} =&
-\sqrt{(l-m+2)(l-m+1)} \, \alpha^+_{l}\,\ket{l+1,m-1}\\
&+ \sqrt{(l+m)(l-m+1)} \, \alpha^0_l \,\ket{l,m-1}\\
&+ \sqrt{(l+m)(l+m-1)}\, \alpha^+_{l-1}\,\ket{l-1,m-1},
\label{rep-Bs}
\end{aligned}
\end{equation}

\begin{equation}
\begin{aligned}
\pi_N(A) \ket{l,m}  =&
-\sqrt{(l-m+1)(l+m+1)} \,\alpha^+_l\,\ket{l+1,m}\\
&+ m \, \alpha^0_l \, \ket{l,m}\\
&+  -\sqrt{(l-m)(l+m)} \,\alpha^+_{l-1}\, \ket{l-1,m}.
\label{rep-A}
\end{aligned}
\end{equation}
where $\alpha^+_l, \alpha^0_l$ are
\begin{equation}
\begin{aligned}
\alpha_0(l) &= \frac{N}{l(l+1)}, \\
\alpha_+(l) &= \sqrt{1 - \frac{N^2}{(l+1)^2}} \frac{1}{\sqrt{(2l+1)(2l+3)}}
\end{aligned}
\end{equation}
\end{pro}

\begin{proof}
The proof is mostly technical and we only sketch here its
main points, especially since parts of it are equivalent
to the Wigner-Eckart theorem on $su(2)$ irreducible
tensors\footnote{I would like to thank F.J. Vanhecke for
raising this point.}.

First, using the equivariance (\ref{equiv}) for $L^3$ we
see that $\pi(A)$ does not change the eigenvalue whereas
$\pi(B),\pi(B^*)$ change it by $\pm 1$. Then, we deduce that for
$X=A,B,B^*$ and for each $l$ we have:
\begin{equation}
\pi(X) (V_l) \subset V_{l-1} \oplus V_l \oplus V_{l+1},
\label{res1}
\end{equation}
where (a priori) we can have multiplicities on the right hand side. To
see that (\ref{res1}) holds, it suffices to study the equivariance
rule for $\rho(L^\pm)^n$ for a suitably chosen $n$. For instance,
if $X=A$, then using the equivariance we verify that:
$$ \rho((L^+)^{l-m+1} \pi(A) \ket{l,m} = 0, \;\;\;
\rho((L^-)^{l+m+1} \pi(A) \ket{l,m} = 0.$$
Therefore $\pi(A) \ket{l,m}$ must be in the intersection of kernels of
$\rho(L^+)^{l-m+1}$ and $\rho(L^-)^{l+m+1}$ with the eigenspace of
$\rho(L^3)$ to the eigenvalue $m$. Therefore the result (\ref{res1}) for
$X=A$ becomes clear.

The equivariance rule for $L^\pm$ allows us to fix
the $m$-dependence of the coefficients in the expansion
(\ref{res1}). We again use the example of $A$ to illustrate
it,
$$
\langle l\!+\!1,m\!+\!2| \rho(L^+) \pi(B) | l, m \rangle  =
\sqrt{l-m}\sqrt{l+m+3}\; \langle l\!+\!1,m\!+\!1 | \pi(B) l,m \rangle, \\
$$
where we used that $(L^+)^* = L^-$. On the other hand, using equivariance:
$$
\begin{aligned}
\langle l\!+\!1,m\!+\!2| \rho(L^+) &\pi(B) | l, m \rangle  =
\langle l\!+\!1,m\!+\!2|  \pi(B) \rho(L^+) | l, m \rangle \\
&= \sqrt{l-m}\sqrt{l+m+1} \; \langle l\!+\!1,m\!+\!2 | \pi(B) l,m\!+\!1 \rangle.
\end{aligned}
$$
By comparing these expressions, we see that
$B_{l,m}^+ = \langle l\!+\!1,m\!+\!1| \pi(B) | l, m \rangle$ must satisfy:
\begin{equation}
B_{l,m}^+ \sqrt{l+m+3} = B_{l,m+1}^+ \sqrt{l+m+1},
\end{equation}
and is has the solution:
$$ B_{l,m}^+ = \sqrt{l+m+1} \sqrt{l+m+2}\;  \alpha^+_l.$$
Similarly we find the other remaining coefficients.

The recurrence relations for $\alpha^+_l$ and $\alpha^0_l$ are
obtained from the sphere radius restriction (\ref{radius}). The
freedom in the choice of the free parameters in their solutions
corresponds to the choice of the smallest possible $|N|$,
whereas the sign of $\alpha^0_l$ fixes the sign of $N$.
\end{proof}

\subsection{Equivariant Spectral Geometry}

To construct the spectral geometry following the axioms \cite{CoRe}, in addition
to the equivariant representation, we need a $\Z_2$-grading $\gamma$ of the
Hilbert space and the reality operator $J$, satisfying $JD = DJ$ and,
\begin{equation}
J \gamma = - \gamma J,  \;\;\;\;\; J^2 =  - 1, \;\;\;\;\; \gamma^2=1, \label{Jgam}
\end{equation}
with the equivariance condition,
\begin{align}
\rho(\ell) \gamma &= \gamma \rho(\ell),\;\; \forall \ell \in \su(2),
\label{gaequi}\\
J \rho(\ell) &= - \rho(\ell) J, \;\; \forall \ell \in \su(2). \label{Jequi}
\end{align}
It follows directly from (\ref{gaequi}) that $\gamma$ must be diagonal
with respect to the chosen basis on $\CH_N$. Additionally, if it commutes
with the representation $\pi_N$ of $\CA(S^2)$ then it must be either
$1$ or $-1$. The Hilbert space with which we construct the spectral
geometry must be a direct sum of at least two spaces $\CH_N$ and $\CH_N'$
with $\gamma$ being $\pm 1$ on them, respectively.

Further, the existence of an antilinear equivariant isometry $J$, fixes
$|N|=|N'|$. This is so, because the commutation rule (\ref{Jgam}) requires
that $J$ is the isometry between $\CH_N$ and $\CH_N'$. Then the
equivariance condition (\ref{Jequi}) requires that $J$ must map
$\ket{l,m}$ to $\ket{l,-m}$, for any $l$, but this is possible only
if $|N|=|N'|$.

It remains to fix the representation (or, in other words, the sign of $N,N'$).
This follows from a further requirement, the commutant condition:
\begin{equation}
J \pi(x) J = \pi(x^*), \;\; \forall x \in \CA(S^2),
\end{equation}
which leads to the solution $\CH= \CH_N \oplus \CH_{-N}$, $N \geq 0$. We shall
denote the basis by $\ket{l,m,\pm} \in \CH_{\pm N}$. So, our spectral data so far
consists of $\CH$, the diagonal representation of $\CA(S^2)$ on it and the
operators $J,\gamma$:
\begin{align}
\gamma \ket{l,m,\pm} &= \pm \ket{l,m,\pm}, \\
J \ket{l,m,\pm} &= i^{2m} \ket{l,-m,\mp}.
\end{align}

Observe, that only for a half-integer $N$ we have the signs of a two-dimensional spectral
geometry, whereas for an integer value of $N$, we have the sign relations corresponding
formally to a six-dimensional (mod $8$) real structure.

\subsection{The Equivariant Dirac operator}

We assume that $D$ is a densely defined symmetric operator, equivariant with
respect to $su(2)$ symmetry, and satisfying the commutation requirements of
spectral geometry in dimension $2$:
$$ DJ = JD, \;\;\;\;\; D\gamma = -\gamma D. $$
It is easy to see that any operator with these properties must
be of the form:
\begin{equation}
D \ket{l,m,\pm} = d(l)^\pm \ket{l,m,\mp},
\end{equation}
where $(d(l)^\pm)^* = d(l)^\mp$ are complex coefficients.

The most important restriction comes from the order one condition, which
in its exact form reads (for commutative geometries),
\begin{equation}
\left[ [D, \pi(x)], \pi(y) \right] = 0, \;\forall x,y \in \CA(S^2).
\end{equation}

\begin{pro}
The order one condition is satisfied exactly only if $N=\oh$
and $d(l) = (l+\oh) d_1$ for any $d_1 \not= 0$.  For the same form
of the Dirac operator but different $N$ the order one condition is
satisfied up to compact operators $\CK(\CH)$:
\begin{equation}
\left[ [D, x], y \right] \in \CK(\CH), \;\forall x,y \in \CA(S^2).
\label{ooco}
\end{equation}
\end{pro}
\begin{proof}
We verify the order one condition on the generators $A,B,B^*$.
First, looking at the coefficient at $\ket{l\!+\!1,m}$ of
$[B^*,[D,B]]\,\ket{l,m}$ we obtain that in order for the to vanish
we must have:
$$ (2l+3)\, d^+(l) - (2l+1) \,d^+(l+1) = 0, $$
which can be satisfied only if $d(l)$ is:
\begin{equation}
d^+(l) = d_1 \cdot (l + \oh). \label{dirac}
\end{equation}
However, putting that back into the expression we obtain that
$[B^*,[D,B]] \ket{l,m,+}$ does not vanish identically but is:
$$
\left[B^*,[D,B]\right] \ket{l,m,+}
= 4 d_1 \frac{(4N^2-1)(2 l^2 + 2l - 1 -2 m^2)}{(2 l-1)(2 l+1)(2 l+3)}
\ket{l,m,-}. $$
Although this does not vanish we can easily see that then
$[B^*,[D,B]]$ is a compact operator. For large $l$:
$$ || \left[B^*,[D,B]\right] \ket{l,m,+} || \leq  2 |d_1| |4N^2-1|
\frac{1}{l}. $$

Similarly, we check other commutators, for instance:
$$
\left[A,[D,B]\right]\ket{l,m,+} = -4 \frac{\sqrt{l-m}\sqrt{l+m+1}(4 N^2-1)
(2m+1)}{(4l^2-1)(2l+3)} \ket{l,m,-}.
$$
and we find that one can always majorize them by $C \frac{1}{l}$, hence
(\ref{ooco}) holds.
\end{proof}

Note that the modified order-one condition does not enforce the coefficient
$\oh$ in the formula (\ref{dirac}) and could be replaced by an arbitrary
coefficient. This freedom in the choice of this component $d_0$ does not
play a role as it is a perturbation of the Dirac operator by its sign.
It appears however, that for the choice we have made, the formulas
simplify significantly.

\subsection{The properties of the quasi-Dirac operators}

We shall briefly discuss the properties of the quasi-Dirac operators
we have found, for an arbitrary $N$. For simplicity we fix $d_1$ to be
$1$. First, observe that apart from the first $2(N\!-\!1)$ eigenvalues, it
has the same spectrum as the standard Dirac operator and therefore
the asymptotic behavior of the eigenvalues is exactly the same.
In particular, $|D|^{-2}$ is Dixmier class and its Dixmier trace is:
\begin{equation}
\hbox{Tr}_\omega |D|^{-2} = 2.
\label{Dtr}
\end{equation}
Clearly, all commutators $[D,x]$, for $x \in \CA(S^2)$ are bounded
operators (it is an easy exercise to verify it), therefore, from the
point of view of the local index calculations, for each $N$ we indeed
have a two-dimensional spectral geometry.

\subsection{Local index calculation}

To get insight into the presented construction we shall explicitly
calculate here the index pairing between the K-homology element
represented by the construction of the spectral geometry and a
chosen element of K-theory using the Connes-Moscovici \cite{CoMo}
local index formula.

Let us recall that having the spectral data we can explicitly assign
to the spectral geometry a cyclic cocycle from the b-B bicomplex.
For the interesting case of a two-cyclic cocycle for the two-dimensional
spectral geometry we have the following even b-B cocycle:
\begin{align}
\phi_0(a_0) &= \tau_{-1}( \gamma \pi(a_0)), \\
\phi_2(a_0,a_1,a_2) &= \frac{1}{2} \tau_0 \left( \gamma
\pi(a_0) [D,\pi(a_1)] [D, \pi(a_2)] |D|^{-2} \right).
\end{align}
where $\tau_q$ is defined as:
$$ \tau_q (T) = \hbox{Res}_{z=0} \hbox{Tr} \left( z^q  T |D|^{-2z} \right).$$
The paring of the above cocycle with the projector $e$ depends only on
the class of $\phi$ and $e$:
$$
\langle [\phi], [e] \rangle =
(2 \pi i)^2 \left( \phi_0(e) - 2 \phi_2(e\!-\!\oh,e,e)\right),$$
where the coefficient is chosen so that the pairing is integral.
Choosing $e$
$$
e = \frac{1}{2} \left( \begin{array}{cc} 1-A & B \\ B^* & 1+A \end{array}\right),
$$
which is a representative of a nontrivial K-theory class of the
sphere we have:
$$
\begin{aligned}
\phi_0(e) &= \phi_0(\pi(e_{11})) + \phi_0(\pi(e_{22})) = \phi_0(1)  \\
&= \tau_{-1}(\gamma) = 0.
\end{aligned}
$$
and
$$
\phi_2(e\!-\!\oh,e,e) = \sum_{i,j,k =1,2} \phi_2(e_{ij}-\oh
\delta_{ij},e_{jk},e_{ki}).
$$
The calculation itself is a technical and rather tedious task.
The crucial point is the following result:
\begin{equation}
\left( \sum_{i,j,k =1,2} (\pi(e_{ij})-\oh \delta_{ij}) [D, \pi(e_{jk})]
[D, \pi(e_{ki})] \right) \ket{l,m,\pm} = \pm N \ket{l,m,\pm}.
\label{Hoch}
\end{equation}
from which we easily get:
\begin{pro}
The pairing between the Chern character of the spectral geometry
$(\CA(S^2), \CH_N \oplus \CH_{-N}, D)$ and the element $e$ of
the $K$-theory is $2N$.
\end{pro}
\begin{proof}
Using (\ref{Hoch}) we explicitly calculate the paring:
$$
\begin{aligned}
(2\pi i)^2 \langle (\phi_0,\phi_2), e \rangle &=
(2\pi i)^2 \left( \phi_0(e) -2 \phi_2(e\!-\!\oh,e,e)\right) \\
&= - (2\pi i)^2 \frac{1}{2} 2N \, \hbox{Res}_{z=0} \hbox{Tr} (|D|^{-2-2z} ) \\
&= - 2 N (2\pi i)^2 \frac{1}{2 (2\pi)^2}  \hbox{Tr}_\omega (|D|^{-2}) \\
&=   4 N (2\pi)^2 \frac{1}{2 (2\pi)^2} = 2N.
\end{aligned}
$$
where we used further the relation between the residue and the Dixmier
trace as well as the result (\ref{Dtr}).
\end{proof}

Observe that as a byproduct of the calculations we have verified the
Hochschild cycle axiom of the spectral triple, as one can easily
see from (\ref{Hoch}).

It appears that the calculation of the explicit formulae for the nonlocal
Connes-Chern character using the Chern character in periodic cyclic cohomology,
for a two-cyclic cocycle is also possible, although it is (obviously) more
complicated. Taking $F=\hbox{sign}(D)$ we have:
$$ \Phi_2(a_0,a_1,a_2) = \frac{1}{4} \hbox{Tr}
\left(\gamma F [F,\pi(a_0)][F,\pi(a_1)][F,\pi(a_2)]\right),$$
After long calculations we obtain:
$$
\begin{aligned}
\hbox{Tr} \left( \sum_{i,j,k =1,2}
\gamma F [F,e_{ij}] [F, e_{jk}] [F, e_{ki}] \right)
&= 4 N^3 \sum_{l=N,N+1, \ldots} \sum_{m=-l}^l \frac{1}{l^2 (l+1)^2} \\
&= 4 N^3 \left(\sum_{l=N}^\infty \frac{2l+1}{l^2(l+1)^2} \right) = 4 N.
\end{aligned}
$$
so the pairing, which could be calculated as $(-2) \Psi_2(e,e,e)$ becomes:
$$ \langle \Phi_2, e \rangle = -2 \Phi_2(e,e,e) = 2 N. $$

\section{Quasi Fermions and Differential Operators}

In the previous section we have demonstrated that the Dirac operator $D_N$ belongs
to a different topological class. Now, we can ask, what would be the physical
properties of fermions whose dynamics would be governed by such {\em quasi-Dirac}.
If we want to pass from $2$-dimensional physics to the real world, it would
be sufficient to extend the spectral geometry in the radial direction. The
appropriate Dirac operator is:
\begin{equation}
D = \gamma \frac{1}{r} \partial_r r + \frac{1}{r} D_N.
\end{equation}

It is easy to see that the only significant difference would be the change in the
eigenvalues, as the spectrum of the quasi-Dirac operator $D_N$ starts with $N$
and not $\frac{1}{2}$. Thus, for a solution of the Dirac equation in a central
potential we shall have the dependence of the energy levels on the angular
momentum slightly different as in the standard case. The lowest angular
momenta up to $N-1$ shall be absent from the energy spectrum. A naive physical
interpretation of that might be that we have a description of fermions with spin
different than $\frac{1}{2}$, thus raising questions of relations with the
Rarita-Schwinger equation for spin $\frac{3}{2}$ particles. The difference
between the two is, however, fundamental as our quasi-Dirac is necessarily
a pseudodifferential operator and therefore the Dirac equation is not
a first order differential equation.

The quasi-fermions might be, similarly as the usual ones, charged, thus
interacting through a $U(1)$ gauge potential. Interestingly, since the gauge
potential correspond to a one-form and these shall not commute with the elements
of the algebra, we shall obtain a mildly noncommutative gauge field theory. This
shall also cause the gauge strength to contain terms quadratic in the gauge
potential, thus leading to self-interactions of the gauge field. Since the order-one
condition is obeyed up to compact operators, in the spectral action the terms shall
appear as next-order correction with respect to the cutoff energy scale.
Nevertheless this mild type of noncommutativity can bring some interesting new
effects, which we shall study in our future work.

The classical equivariant Dirac operator on the Riemannian sphere, which corresponds to
the $N=1$ situation described above, gives a representative of a K-homology class. The
standard method to obtain representant of the remaining classes is to twist the Dirac
spinor bundle $S$, by tensoring it (over the algebra of functions) with a nontrivial
line bundle $E$ over the sphere. Then, the twisted Dirac operator, $D_E$, is associated
to the tensor product connection:
$$ \nabla = \nabla_S \ts 1 + 1 \ts \nabla_E, $$
where $\nabla_S$ is the spin connection and $\nabla_E$ is
a connection on the line bundle $E$. (for the details of construction, explicit
formulae and properties see \cite{JVTok}, \cite{DaSilva} and references therein).

The twisted Dirac operator $D_E$ is again an elliptic, self-adjoint $\Z_2$-graded differential
operator and from the Atiyah-Singer theorem \cite{AS} we know that the index of $D_E^+$ depends
on the Chern number $k$ of the line bundle $E(k)$. Thus, twisting by inequivalent line bundles
we obtain inequivalent Fredholm modules and thus different representants of $K$-homology group
of the sphere. However, we can easily see that the construction does not resemble our case. Indeed,
the Dirac spinor bundle $S$ is a trivial one, constructed as a direct sum of two bundles
of Chern numbers $+1$ and $-1$, $S = E(1) \oplus E(-1)$.
If we tensor $S$ by a line bundle of Chern number $k$, we have:
$$ S \ts E(k) \sim E(k+1) \oplus E(k-1). $$

In our case, however, the generalized spinor bundle was different, as it was $E(N) \oplus E(-N)$, which
is only stably equivalent to a trivial bundle, thus, the quasi-Dirac operator cannot be related to a
twisted Dirac operator. The explicit identification of the basis of the linear space $\CV_N$ (which
we used to construct the spectral triple) with the sections of the line bundle of Chern index $N$
(monopole harmonics) is discussed in \cite{Wu,DaSilva}.

Of course, having the pseudodifferential operator $D_N$, one might look for its principal part,
which should be of order one. This, however, shall necessarily be not invariant under the rotation
group and therefore breaking the rotational symmetry of the system. Hence the conclusion is that (apart from
the case $N=1$ it is possible to construct and describe nontrivial topological sectors of fermions on the
sphere (and therefore also on $\R^3 \setminus \{0\}$) but only using pseudodifferential operators and
slightly extended notion of geometry.

\section{Conclusions}

We have shown that within the framework of noncommutative geometry it is possible to construct
nonequivalent spectral geometries on the two-dimensional sphere. The results open
several possibilities. First of all, it appears that the modification of order-one condition has
more profound consequences than it was believed. In this context it would be nice to answer
the question how the exactness of the order-one condition fixes the choice of the spinor bundle
and the Dirac operator.

To study the analytic properties of the quasi-Dirac operators and to see whether there might by any
differences from the classical case it shall be advisable to find the local (coordinate)
expressions for the corresponding pseudodifferential operators.

Our result has also some implications for noncommutative geometries. We have shown that for the two-sphere
from each of different K-homology classes it is possible to obtain unbounded Fredholm modules, with the
order-one axiom satisfied up to compact operators. These are the $q=1$ limits of Dirac operators studied
for quantum Podles spheres \cite{Sitarz-quasi} and are in the noncommutative worlds not easily
distinguished from the deformation of the usual spectral geometry.

{\bf Acknowledgements:}
Part of the work was used for a EU project
``Researchers in Europe Initiative 2005'',
FP6-2004-Mobility-13, Proposal 019697
assigned to Guy Barrett and Andrew Elliott from
St Robert of Newminister Catholic School and Sixth
Form College, Durham, UK.

The author would like to thank the Hausdorff Research Institute  for Mathematics
for hospitality and support.

\end{document}